\documentclass[10pt,twoside,twocolumn,english,aps,manuscript,aps,preprint,showpacs,superscriptaddress,showkeys]{revtex4}
\usepackage{lmodern}
\usepackage[T1]{fontenc}
\usepackage[latin9]{inputenc}
\usepackage[active]{srcltx}
\usepackage{textcomp}
\usepackage{graphicx}
\usepackage{setspace}

\makeatletter
\@ifundefined{textcolor}{}
{%
 \definecolor{BLACK}{gray}{0}
 \definecolor{WHITE}{gray}{1}
 \definecolor{RED}{rgb}{1,0,0}
 \definecolor{GREEN}{rgb}{0,1,0}
 \definecolor{BLUE}{rgb}{0,0,1}
 \definecolor{CYAN}{cmyk}{1,0,0,0}
 \definecolor{MAGENTA}{cmyk}{0,1,0,0}
 \definecolor{YELLOW}{cmyk}{0,0,1,0}
}

\makeatother

\usepackage{babel}
\begin{document}

\title{Evidence of surface superconductivity and multi\,-\,quanta vortex
states in a weakly\,-\,pinned single crystal of Ca$_{3}$Ir$_{4}$Sn$_{13}$}

\author{Santosh Kumar}

\email{santoshkumar@phy.iitb.ac.in}

\selectlanguage{english}%

\address{Department of Physics, Indian Institute of Technology Bombay, Mumbai
400076, India}

\author{Ravi P. Singh}

\altaffiliation{Present address: Department of Physics, Indian Institute of Science Education and Research, Bhopal 462066, India.}

\selectlanguage{english}%

\address{Department of Condensed Matter Physics and Materials Science, Tata
Institute of Fundamental Research, Mumbai 400005, India.}

\author{A. Thamizhavel}

\address{Department of Condensed Matter Physics and Materials Science, Tata
Institute of Fundamental Research, Mumbai 400005, India.}

\author{C. V. Tomy}

\address{Department of Physics, Indian Institute of Technology Bombay, Mumbai
400076, India}

\author{A. K. Grover}

\address{Department of Condensed Matter Physics and Materials Science, Tata
Institute of Fundamental Research, Mumbai 400005, India.}

\address{Department of Physics, Panjab University, Chandigarh 160014, India.}
\begin{abstract}
We report here the observation of anomalous paramagnetic signal(s)
in the isofield field\,-\,cooled cool\,-\,down magnetization scans
($M_{FCC}(T)$) recorded for a single crystal of a low $T_{c}$ superconductor
Ca$_{3}$Ir$_{4}$Sn$_{13}$. Novel features emanating from the $M_{FCC}(T)$
response include an oscillatory magnetization behaviour below $T_{c}$
and a rich multiplicity (non\,-\,uniqueness) in magnetization ranging
from diamagnetism to paramagnetism at a given $H$, $T$ value. The
metastability in $M_{FCC}(T)$ has been ascribed to non\,-\,unique
coexistence of multi\,-\,quanta vortex states ($L\Phi_{0}$, $\Phi_{0}=hc/2e$,
$L>1$) and single quantum ($L=1$, Abrikosov) vortices. Additionally,
the isothermal $M(H)$ scans recorded across a short window of temperature
just below $T_{c}$ show evidence for only the multi\,-\,quanta
vortex states in the domain of surface superconductivity, with no
fingerprint(s) of pinned Abrikosov lattice.
\end{abstract}



\maketitle

\section{Introduction}

In recent years, the exploration of temperature variation of the field\,-\,cooled
cool\,-\,down (FCC) magnetization ($M_{FCC}(T)$) measurements at
low fields in a variety of superconductors \cite{key-1,key-2,key-3,key-4,key-5,key-6,key-7,key-8,key-9,key-10,key-11,key-12}
have revealed characteristics which have been ascribed to the phenomenon
of surface superconductivity \cite{key-13}. Many interesting features
reported in the $M_{FCC}(T)$ responses at different fields in a weakly\,-\,pinned
single crystal of a low $T_{c}$ ($\sim8.3$\,K) superconductor,
Ca$_{3}$Rh$_{4}$Sn$_{13}$ \cite{key-10}, include the occurrence
of anomalous paramagnetic magnetization \textit{a} \textit{la} paramagnetic
Meissner effect (PME), modulations overriding PME like response, intersection
of the $M_{FCC}(T)$ curves for different field values across a field
interval at a characteristic temperature $T_{VL}^{*}$, below the
onset temperature, $T_{c}$, of superconducting response, etc. Investigations
in the low field regime in Ca$_{3}$Rh$_{4}$Sn$_{13}$ \cite{key-10}
had also revealed a concave curvature in the temperature dependence
of the upper critical field line ($H_{c2}(T)$, $T_{c}(H)$) near
its $T_{c}(0)$, which was conjectured to reflect the persistence
of superconductivity at the surface \cite{key-13} at low fields,
up to third critical field $H_{c3}$ values, which were well beyond
the notional second critical fields, $H_{c2}$. The $T_{c}(H)$ values
at low field were thus argued to represent the temperatures at which
the superconductivity nucleated only on the surface, whereas the temperature
$T_{VL}^{*}$ was identified as the temperature below which the superconductivity
permeated into the interior of the single crystal of Ca$_{3}$Rh$_{4}$Sn$_{13}$
\cite{key-10}.

We have recently explored \cite{key-14} the vortex phase diagram
in a single crystal of another low $T_{c}$ superconducting compound,
Ca$_{3}$Ir$_{4}$Sn$_{13}$, which is isostructural to Ca$_{3}$Rh$_{4}$Sn$_{13}$
\cite{key-10,key-15}. The dc magnetization data ($M(H)/M(T)$) in
Ca$_{3}$Ir$_{4}$Sn$_{13}$ \cite{key-14} revealed a long tail surviving
beyond the irreversibility field/temperature ($H_{irr}$\,/\,$T_{irr}$),
in contrast to the anticipated linear variation of $M(H)$ (or $M(T)$)
as a function of $H$ (or $T$) up to the end of the superconducting
region. The experimental values of the upper critical field/superconducting
transition temperature ($H_{c2}$\,/\,$T_{c}$) in Ca$_{3}$Ir$_{4}$Sn$_{13}$
were found to be significantly higher than those which can be ascertained
following the mean field description \cite{key-16} of equilibrium
magnetization in a type\,-\,II superconductor. A field\,-\,temperature
line, viz., ($H^{*}$,$T^{*}$), was drawn in the vortex phase diagram
of Ca$_{3}$Ir$_{4}$Sn$_{13}$ \cite{key-14}, which runs parallel
to the $H_{c2}(T)$ line. The two field-temperature lines, ($H^{*}$,$T^{*}$)
and $H_{c2}(T)$, intersect the temperature axis at $T\approx6.85$\,K
and $T_{c}\approx7.1$\,K, respectively in Ca$_{3}$Ir$_{4}$Sn$_{13}$.
In an attempt to comprehend the underlying physics prevailing in the
region bounded between the ($H^{*}$,$T^{*}$) and ($H_{c2}$,$T_{c}$)
lines in the vortex phase diagram of Ca$_{3}$Ir$_{4}$Sn$_{13}$
{[}see Fig. 7, ahead{]}, we now present new results of dc magnetization
measurements in the said region in this compound. We have observed
anomalous paramagnetic signals, below $T_{c}$, in the field\,-\,cooled
cool\,-\,down magnetization $M_{FCC}(T)$ measurements. These paramagnetic
signals display path dependence, i.e., at a given ($H$,$T$) value,
one can observe multiplicity in response. The paramagnetic peaks in
$M(T)$ curves also imbibe modulations. The said features in $M_{FCC}(T)$
behaviour observed in the current investigations can be ascribed to
the occurrence of multi\,-\,quanta ($L\Phi_{0}$, $\Phi_{0}=hc/2e$,
$L>1$) vortex states at the onset of surface superconductivity and
various transitions amongst multi\,-\,quanta states before the emergence
of (single quantum, $L=1$) Abrikosov flux lines. The nature of isothermal
M\,-\,H scans in the close proximity of $T_{c}$ lend further support
to the notion of multi\,-\,quanta states in the domain of surface
superconductivity in the crystal of Ca$_{3}$Ir$_{4}$Sn$_{13}$ under
study.

\section{Experimental details}

The single crystal specimen chosen for the present investigation is
the same that was employed in Ref. \cite{key-14}. The direction of
magnetic field was maintained parallel to the plane of the platelet
shaped sample, i.e., normal to its smallest dimension (i.e., thickness).
The amplitude of vibration of the sample for the dc magnetization
measurements performed using a vibrating sample magnetometer (VSM)
was kept low ($=0.5$\,mm) for most of the runs. The magnetization
was recorded in the same instrument as utilized in Ref. \cite{key-14},
viz., the Superconducting Quantum Interference Device Vibrating Sample
Magnetometer (SQUID\,-\,VSM, Quantum Design Inc., USA). 

\begin{figure}[!t]
\begin{centering}
\includegraphics[scale=0.33]{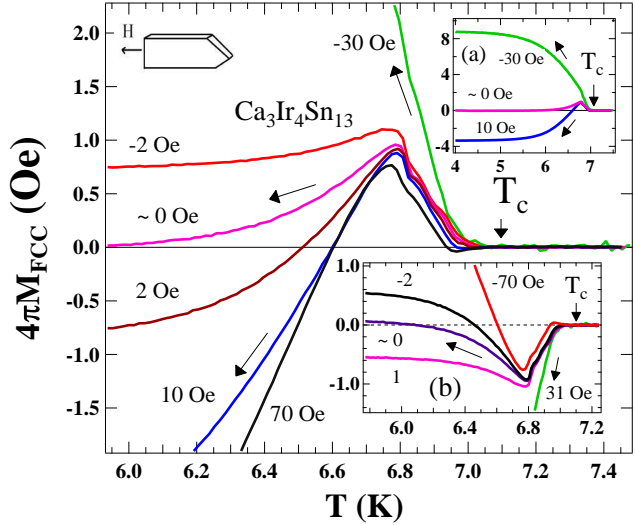}
\par\end{centering}

\protect\caption{{\footnotesize{}(Color online) Portions of $M_{FCC}(T)$ curves near
$T_{c}$ obtained at various magnetic field values, as indicated.
Magnetic field was applied parallel to plane of the crystal. Anomalous
paramagnetic signals can be noticed in the $M_{FCC}(T)$ curves for
$H>0$. The inset (a) illustrates $M_{FCC}(T)$ responses at $H=-30$\,Oe,
$0$\,Oe and $10$\,Oe in the range $4$\,K\,$<T<7.5$\,K. The
inset (b) shows the $M_{FCC}(T)$ responses recorded with a different
history of the magnet (see text for details).}}
\end{figure}

\section{Results}

\subsection{Paramagnetic behaviour and the anomalous features in field-cooled
magnetization measurements}

Figure~1 illustrates the temperature dependences of field\,-\,cooled
cool\,-\,down magnetization responses, $M_{FCC}(T)$, for various
fields ($\mid H\mid<100$\,Oe), as indicated. Magnetization values
were recorded while cooling the crystal from a temperature $T>T_{c}$
in the presence of a chosen value of applied magnetic field. Considering
the presence of a remnant field appearing (when the current is set
to zero) in the superconducting magnet of the SQUID\,-\,VSM, we
initially attempted to estimate its magnitude and polarity by a procedure
described ahead. Starting from a higher (say) positive value, we set
the current in the superconducting magnet to zero. We anticipate a
remnant field in the superconducting magnet, whose sign and value
depend on the history of current in the magnet. We start the process
of recording the magnetization data by selecting a given remnant field,
whose precise value, we do not know apriori. We noticed that the $M_{FCC}(T)$
curve in the superconducting state remained positive down to the lowest
temperature for $T<T_{c}$, which indicated that the starting remnant
field was of negative polarity. We then progressively increased the
applied magnetic field towards higher positive values by incrementing
magnet current in small steps, and measured $M_{FCC}(T)$ responses
at each selected field value. It was found that the saturated value
of the $M_{FCC}(T)$ curves (see the nearly temperature\,-\,independent
$M_{FCC}$ values below about $5.5$\,K in the inset panel (a) of
Fig.~1) crossed over to the negative side after displaying an anomalous
(paramagnetic) positive peak, just below $T_{c}$ (cf. main panel
of Fig.~1), when the set value of applied field in magnet was $\geq+30$\,Oe.
We thus reckoned that the initial remnant field in the magnet was
near $-30$ Oe in the series of $M_{FCC}$ measurements performed,
some of which are shown in Fig.~1. Keeping this estimate of remnant
field in view, we arrived at the effective values of the magnetic
field corresponding to each of the $M_{FCC}(T)$ curves in Fig.~1.
Note that the $M_{FCC}(T)$ response for $H=-30$\,Oe (cf. inset
panel (a) of Fig.~1) displays a usual (i.e., diamagnetic behaviour)
superconducting transition occurring at about $T=7.1$\,K (i.e.,
there is no apparent anomalous \textquoteleft paramagnetic\textquoteright{}
magnetization feature at $H=-30$\,Oe). However, on enhancing the
field values from negative to positive side gradually, an anomalous
positive (i.e., paramagnetic) magnetization just below $T_{c}$ becomes
apparent and such a behaviour was seen to persist for higher positive
fields upto about $H=1$\,kOe (the magnetization curve for $1$\,kOe
is though not displayed in Fig.~1). We also checked for the presence
of the anomalous (paramagnetic) feature if the applied field is increased
to higher negative values, starting from the negative value of the
remanant field of the magnet. We noted that the anomalous PME like
feature was not present for large negative fields. It is emphasised
that the nomenclature \textquoteleft positive\textquoteright{} or
\textquoteleft negative\textquoteright{} for field is arbitrary; the
so called positive and negative signs of the field produced by the
magnet are just the phase reversed signs. To satisfy ourselves, we
repeated the record of the entire $M_{FCC}(T)$ data by beginning
the process of measurements, on obtaining remanant field of positive
polarity instead of negative value in the main panel of Fig.~1. 

To reiterate, after setting the current (in the magnet) once again
to zero from the higher negative values (instead of higher positive
values as earlier), we recorded the $M_{FCC}(T)$ data, which remained
negative (i.e., diamagnetic) for $T\ll T_{c}$, and confirmed the
remnant field to be of positive polarity. Thereafter, we gradually
incremented the field from notional zero value (i.e., remenant field)
to higher negative values and recorded the $M_{FCC}(T)$ data at each
field value. The $M_{FCC}(T)$ responses now show the behavior, which
is (nearly) a mirror image of those obtained earlier with the magnet
having initial remnant field of negative polarity (see inset panel
(b) of Fig.~1). The anomalous (negative) peak in $M_{FCC}(T)$ curves
can now be noticed only for the negative fields, whereas no such peak
(i.e., paramagnetic) feature could be observed for the positive fields.
Here also, the paramagnetic nature is not seen if the applied magnetic
field is increased to higher positive values. It is now instructive
to note that irrespective of the magnet history, the $M_{FCC}(T)$
response initially displays a usual diamagnetic transition across
the superconducting transition (cf. $M_{FCC}(T)$ curve for $H=-30$\,Oe
in the main panel and the inset panel (a) of Fig.~1, and for $H=31$
Oe in the inset panel (b)) for zero current in the magnet. However,
gradually reversing the sign of the field, from positive to negative
or \textit{vice-versa}, results in the above mentioned anomalous features
in $M_{FCC}(T)$ plots. We are thus lead to surmise that there is
no difference between the positive and negative fields in the context
of the anomalous peak feature as it can be reproduced for both the
signs of the field, by simply reversing the initial magnet condition
(i.e., the sign of the remnant field). Such an exercise had not been
carried out while reporting similar results in Ca$_{3}$Rh$_{4}$Sn$_{13}$
in Ref.~\cite{key-10} and perhaps had led to an erroneous surmise
as if the sign of earth\textquoteright s field had some influence
in determining the asymmetry between two signs of the applied field. 

It is worthwhile to explore the possibility of encountering the anomalous
paramagnetic feature in $M_{FCC}(T)$ curves (Fig.~1) in some other
circumstances, say, via further change(s) in the experimental conditions.
Usually, the magnitude of the remnant field gets reduced significantly
by reducing the current in the magnet to zero value in an oscillating
mode option of the current source. Following this alternative, the
magnet current was sequentially reduced to zero in an oscillating
mode from a high value of positive field: $70$\,kOe \textrightarrow{}
$10$\,kOe \textrightarrow{} $1$\,kOe \textrightarrow{} $0$\,Oe
(we designate this condition of remanant field as the remanant state\,-\,A).
We then recorded the $M_{FCC}(T)$ responses as shown in Fig.~2(a).
The remnant field in the state\,-\,A was independently reckoned
to be about $-5$\,Oe. The applied magnetic field was then increased
from this remanant value to higher positive values in small steps.
The paramagnetic peak in magnetization could be observed for a few
field values in close proximity of the zero field, however, the peak
height is substantially suppressed as compared to the corresponding
values in Fig.~1. We then reversed the initial state of the magnet
by setting the current such that the reduction in field in an oscillating
mode goes through the sequence: $-70$\,kOe \textrightarrow{} $-10$\,kOe
\textrightarrow{} $-1$\,kOe \textrightarrow{} $0$\,Oe (designated
as the remanant state\,-\,B). The remnant field was found to be
$\sim+3$\,Oe on this occasion. $M_{FCC}(T)$ curves were then recorded
(cf. Fig.~2(b)) by increasing the current from zero value to higher
negative values in small steps. It is apparent that the behavior of
$M_{FCC}(T)$ curves in Fig.~2(b) almost echo the observations in
Fig.~2(a) (for remnant field of negative polarity in state\,-\,A).
\begin{figure}[!t]
\begin{centering}
\includegraphics[scale=0.42]{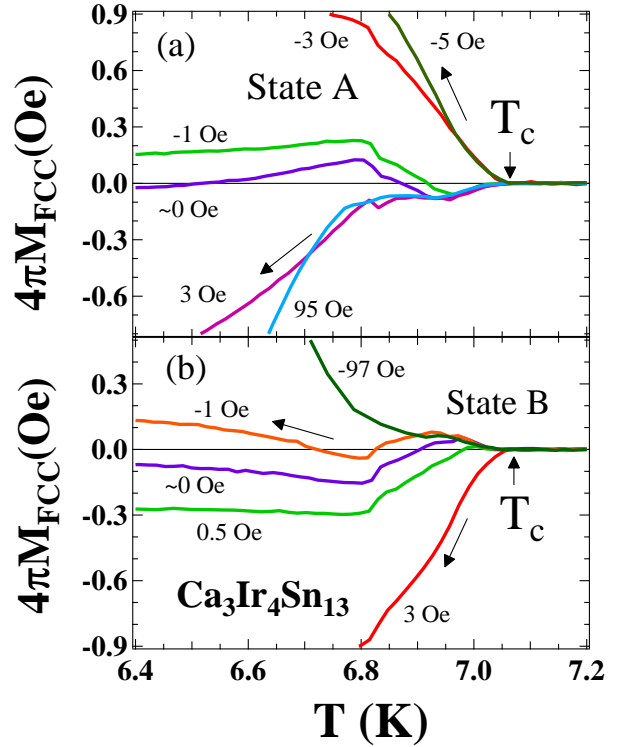}
\par\end{centering}

\protect\caption{{\footnotesize{}(Color online) $M_{FCC}(T)$ responses at various
fields with the initial condition of the magnet in (a) state\,-\,A
(achieved by oscillating the field and creating a small value of remnant
field of negative polarity) and (b) state\,-\,B (oscillating the
field and creating a small value of remnant field with positive polarity).
The details of the two remanant states (A and B) have been explained
in the text.}}
\end{figure}
 
\begin{figure}[!t]
\begin{centering}
\includegraphics[scale=0.4]{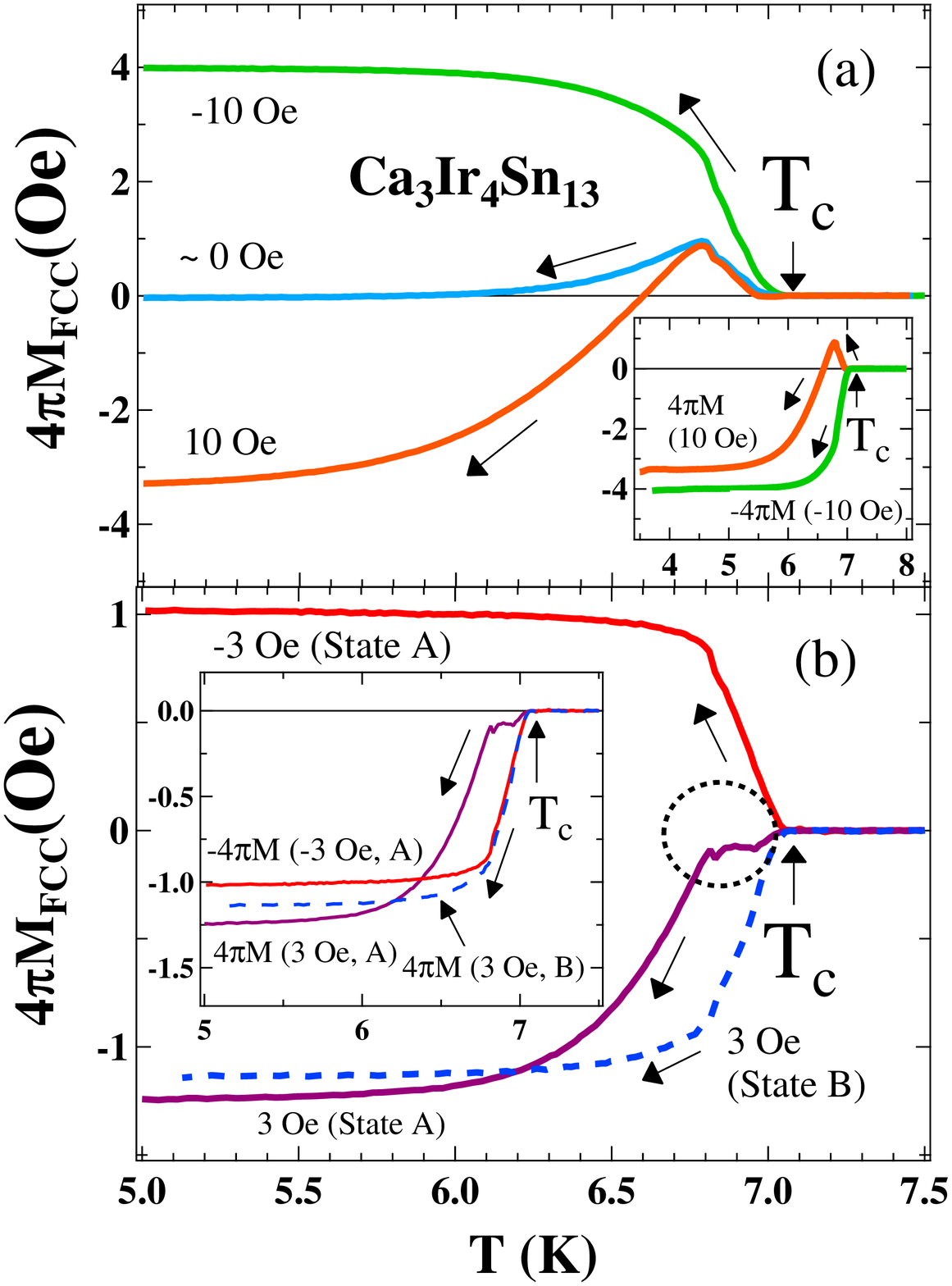}
\par\end{centering}

\protect\caption{{\footnotesize{}(Color online) (a) $M_{FCC}(T)$ for $H=0$\,Oe,
$\pm10$\,Oe. The inset panel in Fig.~3(a) compares an inverted
$M_{FCC}(T)$ (i.e., $-4\pi M$\,($-10$\,Oe)) curve at $H=-10$\,Oe
with $M_{FCC}(T)$ at $H=10$\,Oe; the two curves do not overlap.
(b) $M_{FCC}(T)$ at $H=\pm3$ Oe (state\,-\,A) and at $H=+3$\,Oe
(state\,-\,B). The inset panel in Fig.~3(b) shows an inverted $M_{FCC}(T)$
at $H=-3$\,Oe (state\,-\,A) compared with $M_{FCC}(T)$ at $H=3$\,Oe
of states A and B.}}
\end{figure}

\subsection{Metastability and non\,-\,uniqueness in $M_{FCC}$ response}

Figure~3 shows two $M_{FCC}(T)$ curves obtained for the same field
value, but with opposite polarities, set in the magnet of SQUID\,-\,VSM.
The panel (a) in Fig.~3 includes $M_{FCC}(T)$ for $H=\pm10$\,Oe
extracted from the main panel of Fig.~1. The panel (b) in Fig.~3
contains $M_{FCC}(T)$ curves for $H=\pm3$\,Oe recorded in the state\,-\,A
acquired from Fig.~2(a) and for $H=+3$\,Oe obtained in the state\,-\,B
from Fig.~2(b). It is well apparent from Fig.~3 that the saturated
values of the $M_{FCC}(T)$ below $T\approx5.5$\,K are different
for $\mid H\mid=10$\,Oe (cf. main panel in Fig.~3(a)) as well as
for $\mid H\mid=3$\,Oe (cf. main panel in Fig.~3(b)). To elaborate
this further, we show in the inset panel of Fig.~3(a), a comparison
of $M_{FCC}(T)$ for $H=+10$\,Oe and an inverted $M_{FCC}(T)$ curve
for $H=-10$\,Oe (inverted about the $T$\,-\,axis, i.e., $-M$\,($H=-10$\,Oe)).
These two curves, in the inset panel were expected to overlap, however,
they showed significant differences, with a paramagnetic peak present
in one case below $T_{c}$. Similar non\,-\,uniqueness in magnetization
can also be observed at $\mid H\mid=3$ Oe (see Fig.~3(b). In the
inset panel of Fig.~3(b), we show an inverted $M_{FCC}(T)$ curve
(see dashed curve) at $H=-3$\,Oe (i.e., $-4\pi M$\,($-3$\,Oe,\,A))
obtained in the state\,-\,A, which nearly overlaps with $M_{FCC}(T)$
at $H=3$\,Oe of state\,-\,B for a short window of temperature
below $T_{c}$. However, the saturated $M_{FCC}(T)$ values (below
about $6$\,K) for all three curves in the inset panel of Fig.~3(b)
can be seen to be different. In addition, the $M_{FCC}(T)$ at $H=3$\,Oe
obtained in the state\,-\,A also exibits undulations just below
$T_{c}$ which is well depicted by the encircled portion of this curve
in the main panel of Fig.~3(b). However, such large undulations are
not prominently seen in the other two curves ($-4\pi M$\,($-3$\,Oe,\,A)
and $4\pi M$\,($3$\,Oe,\,B)) in Fig.~3(b). 

The curious metastability and non\,-\,uniqueness observed in $M_{FCC}(T)$
data (Fig.~3) motivated us to seek information on the homogeneity
of the magnetic field inside the superconducting magnet. A superconducting
specimen moving in an inhomogeneous field can yield erroneous magnetization
values \cite{key-17}, as the magnetization of the sample changes
to counter the inhomogeneity in the field while moving. We decided
to record the field profiles (i.e., field variation vs distance traversed
on the axis of the magnet), using the flux gate option in SQUID\,-\,VSM
of Quantum Design Inc., USA. Since the said flux gate option does
not operate in an environment where the field value is more than $\mid10\mid$\,Oe,
the field profiles for remnant states obtained in Fig.~1 could not
be traced. Hence, we prepared the magnet in the states A and B (where
remanant field magnitude is expected to be less than $10$\,Oe).
We recorded the field profiles (Fig.~4(a)) in these two states (where
the remnant fields are negative and positive, respectively) as a function
of distance ($r$) from the centre of the gradiometer coil of SQUID\,-\,VSM.
We have also plotted in the inset panel of Fig.~4(a), the (parameterized)
field inhomogeneity, $\triangle H/H(0)$ vs $r$, where $\triangle H$\,($=H(r)-H(0)$),
is the difference between the field value at a given distance, $r$
from the field value at the center of the magnet. The field inhomogeneity
experienced by the sample over a scan length of $2.5$\,mm is found
to be negligibly small ($<10^{-2}$\,Oe), as shown in the inset of
Fig.~4(a). It is important to note that we had maintained the amplitude
of sample vibration to be $\approx0.5$\,mm for the curves plotted
in Figs.~1 to 3. In addition, even for higher amplitudes ($\approx4$\,mm),
the maximum value of the field inhomegeneity ($\triangle H/H(0)$)
is about $20$\,mOe. This prompted us to record the $M_{FCC}(T)$
curves by varying the amplitudes of vibration in a chosen fixed field
($=8$\,Oe in Fig.~4(b)). The inset in Fig.~4(b) displays a comparison
of the $M_{FCC}(T)$ curves recorded for the amplitudes of $0.5$\,mm
and $5$\,mm in $H=8$\,Oe. It was interesting to find that paramagnetic
peak was evident for the smallest chosen amplitude of $0.5$\,mm.
On the other hand, the $M_{FCC}(T)$ remained diamagnetic for $5$\,mm
amplitude in the entire temperature range, $T<T_{c}$. The $M_{FCC}(T)$
response in the inset panel of Fig.~4(b) shows a path dependence
which resembles closely with the non\,-\,unique (path\,-\,dependent)
$M_{FCC}(T)$ data shown in the inset panel of Fig.~3(a). 

\begin{figure}[!t]
\begin{centering}
\includegraphics[scale=0.43]{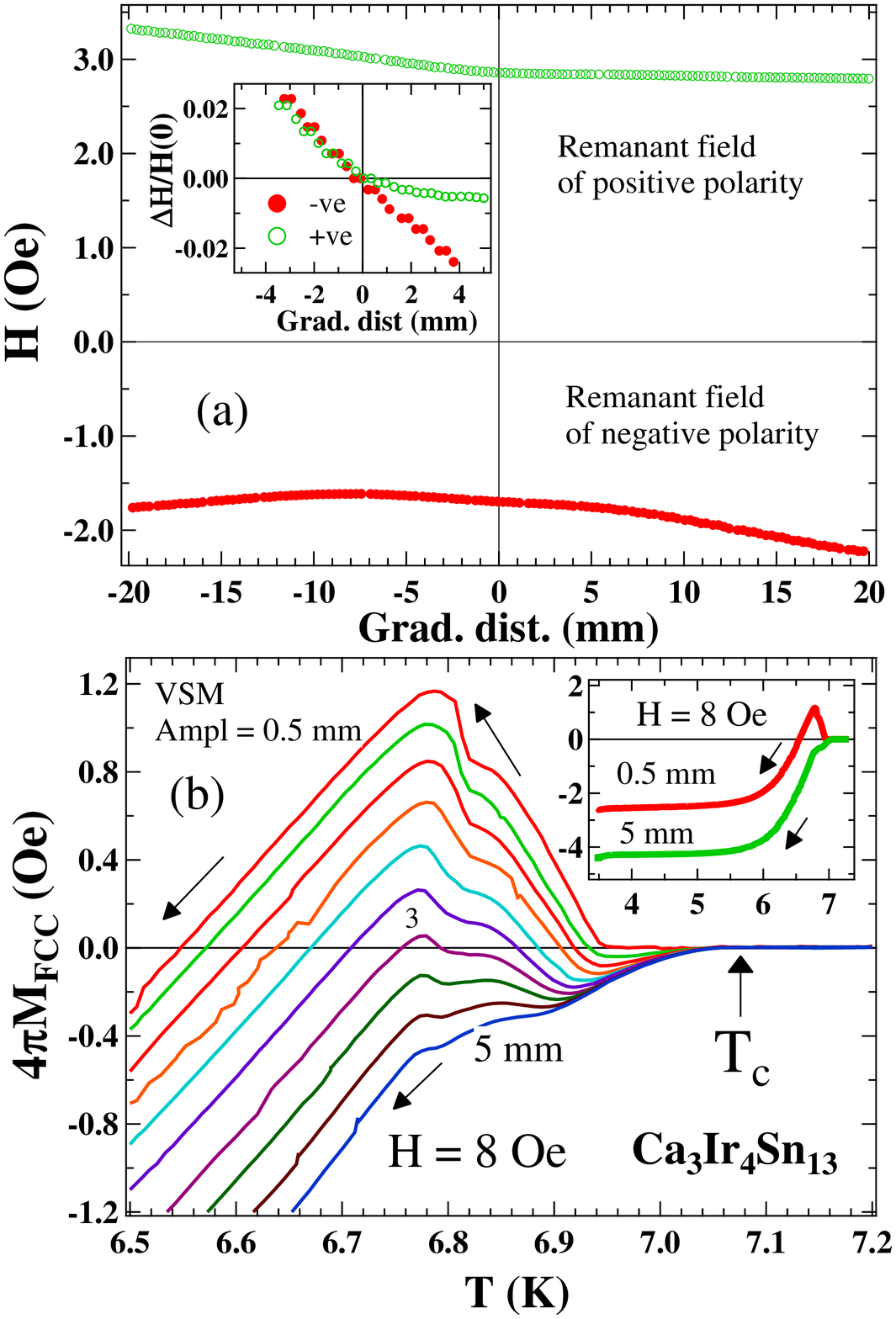}
\par\end{centering}

\protect\caption{{\footnotesize{}(Color online) (a) Remnant field profiles for state\,-\,A
and state\,-\,B recorded against the distance $r$ from centre of
the gradiometer detection coil in the SQUID\,-\,VSM, Quantum Design
Inc., using the flux gate option. An inset in Fig.~4(a) shows a measure
of field inhomogeneity, $\triangle H/H(0)$, where $\triangle H=H(r)-H(0)$,
with distance $r$ for both the states A and B. (b) $M_{FCC}(T)$
responses in the vicinity of $T_{c}$ recorded at different amplitudes
of vibration (varying from $0.5$\,mm to $5$\,mm) in $H=8$\,Oe.
Magnetization below $T_{c}$ is path dependent and exhibits wide variations
from paramagnetic to diamagnetic values. The inset in Fig.~4(b) displays
$M_{FCC}(T)$ in the range, $3.5$\,K\,$<T<T_{c}$ for the two extreme
values of the amplitude of vibration, $0.5$ and $5$\,mm.}}
\end{figure}
The main panel of Fig.~4(b) shows the evolution of the paramagnetic
peak behavior when the amplitude of vibration was progressively incremented
from $0.5$\,mm to $5$\,mm in steps of $0.5$\,mm. All the $M_{FCC}(T)$
curves traverse different paths (below $T_{c}$) as the amplitude
changes. Also, the magnetization response can be seen to be oscillating
in the range $6.8$\,K\,$<T<T_{c}$, as apparent from Fig.~4(b).
It is pertinent to note that the path dependence in $M_{FCC}(T)$
(inset panel of Fig.~4(b)) is not an attribute of the field inhomogeneity
in the magnet of SQUID\,-\,VSM, as we have already examined in Fig.~4(a)
that the $\triangle H/H(0)$ values over scan length of up to $4$\,mm
are not sufficient to change the $M$ values so drastically below
the $T_{c}$. In addition, we also checked for any possible artefact
that could arise in SQUID\,-\,VSM instrument by repeating measurements
on a standard superconducting Indium sample and measured its magnetization
response at a fixed chosen field in different amplitudes (from $0.5$
to $8$\,mm) below $T_{c}$. We found no change in the magnetization
values in $M_{FCC}(T)$ scans with the change in amplitude in the
case of Indium sample. This assured us that the results in Fig.~4
in the case of Ca$_{3}$Ir$_{4}$Sn$_{13}$ need comprehension in
terms of physics of superconductivity.
\begin{figure}[!t]
\begin{spacing}{0.84999999999999998}
\begin{centering}
\includegraphics[scale=0.43]{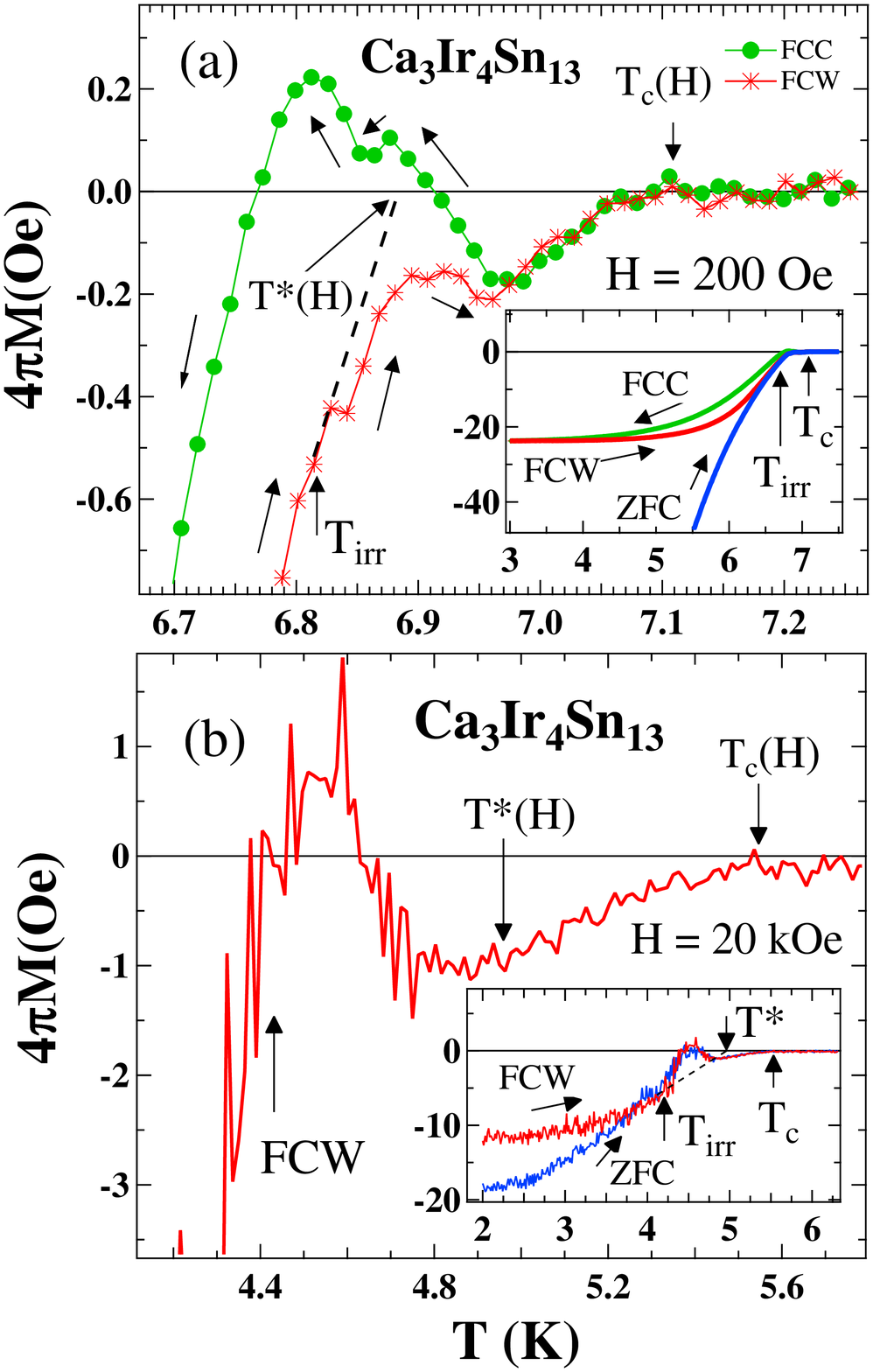}
\par\end{centering}
\end{spacing}

\begin{spacing}{0.84999999999999998}
\protect\caption{{\footnotesize{}(Color online) (a) Expanded portions of $M_{FCW}(T)$
and $M_{FCC}(T)$ curves obtained near $T_{c}$ in $H=200$\,Oe.
A linear extrapolation (dashed line) of magnetization beyond the merger
(at $T_{irr}$) of $M_{ZFC}(T)$ and $M_{FCW}(T)$ curves identifies
a characteristic temperature $T^{*}(H)$, while $T_{c}(H)$ marks
the onset of diamagnetism. A characteristic oscillatory behavior in
$M(T)$ can be noticed below $T_{c}(H)$. An inset in Fig.~5(a) displays
the magnetization response in all the three modes in the temperature
range $3$\,K\,$<T<T_{c}$. (b) $M_{FCW}(T)$ responses focusing
attention on the oscillatory characteristic below $T_{c}(H)$ in $H=20$\,kOe.
The inset in Fig.~5(b) shows the $M_{ZFC}(T)$ and $M_{FCW}(T)$
in $H=20$\,kOe in the temperature range, $2$\,K\,$<T<6$\,K.}}
\end{spacing}
\end{figure}

\subsection{Oscillatory magnetization response in ($M(T)$) scans at higher fields}

The inset panel of Fig.~5(a) shows the temperature variations of
the zero field\,-\,cooled ($M_{ZFC}$), the field\,-\,cooled warm\,-\,up
($M_{FCW}$) and the field\,-\,cooled cool\,-\,down ($M_{FCC}$)
magnetization responses at $H=200$\,Oe. The $M_{FCC}(T)$ and $M_{FCW}(T)$
curves, which nearly overlap at lower temperatures ($T<3$\,K), get
separated at higher temperatures. The $M_{FCW}(T)$ and the $M_{ZFC}(T)$
curves can be seen to merge at a temperature, identified and marked
as the irreversibility temperature, $T_{irr}$ in the main panel of
Fig.~5(a). According to a mean field description of type\,-\,II
superconductors {[}16{]}, the linear extrapolation of the reversible
(i.e., equilibrium) magnetization above $T_{irr}$ is expected to
yield the superconducting transition temperature, $T_{c}(H)$. However,
in the present case, the linear extrapolation of $M(T)$ curve above
$T_{irr}$, identifies a characteristic temperature, $T^{*}(H)$,
which is found to be located significantly lower than the observed
superconducting transition temperature, $T_{c}(H)$, marking the onset
of diamagnetism. The $M_{FCC}(T)$ and $M_{FCW}(T)$ curves exhibit
an unusual oscillatory behavior in the vicinity of $T^{*}(H)$. $M_{FCC}(T)$
curve also displays paramagnetic magnetization persisting well below
$T^{*}(H)$. The oscillatory feature is not just restricted to $M(T)$
curves at low fields, but it can also be observed at higher fields.
The inset panel in Fig.~5(b) illustrates the $M_{ZFC}(T)$ and $M_{FCW}(T)$
curves recorded in $H=20$\,kOe. The irreversibility temperature,
$T_{irr}$ and the characteristic temperature, $T^{*}(H)$ are identified
and marked in the inset of Fig.~5(b). An expanded plot of $M_{FCW}(T)$
in the vicinity of $T_{c}$ (main panel of Fig.~5(b)) shows the oscillatory
magnetization response in $H=20$\,kOe. In addition, the $M_{FCW}(T)$
curve in $H=20$\,kOe is also paramagnetic below $T^{*}(H)$, like
the $M_{FCC}(T)$ curve in $H=200$\,Oe, as is apparent from a comparison
of respective curves in the main panels of Fig.~5(a) and 5(b).

\subsection{Metastable magnetization response and evidence for multi\,-\,quanta
($L>1$) states}

\begin{figure}[!t]
\begin{centering}
\includegraphics[scale=0.43]{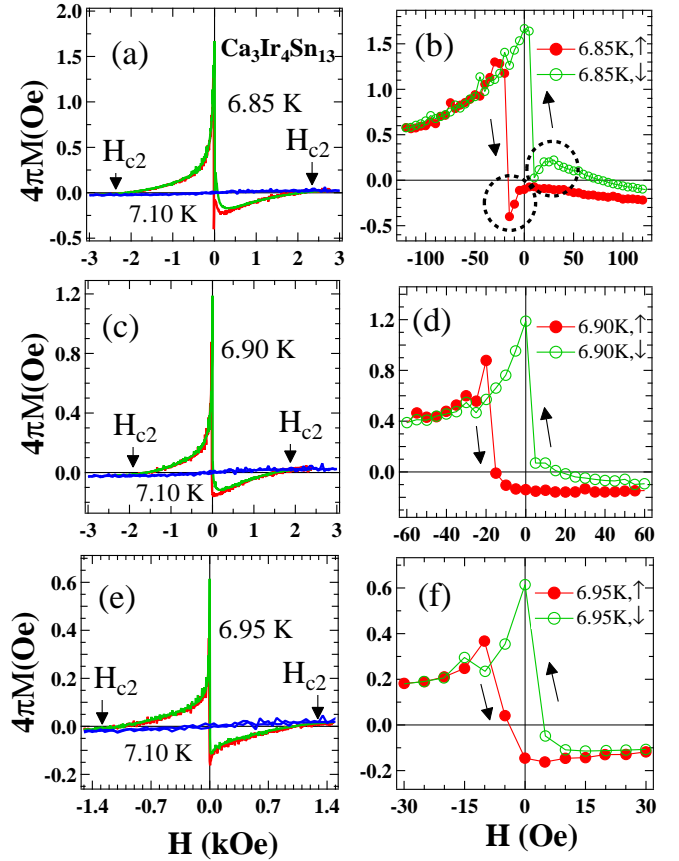}
\par\end{centering}

\protect\caption{{\footnotesize{}(Color online) Isothermal $M(H)$ scans recorded at
(a) $T=6.85$\,K, (c) $T=6.90$\,K and (e) $T=6.95$\,K. $M(H)$
curve at $T=7.1$\,K (normal state) is also shown in all three panels.
Expanded portions of $M(H)$ curves near $H=0$ are displayed in the
right panels, (b) ($6.85$\,K), (d) ($6.90$\,K) and (f) ($6.95$\,K).
Features near zero field, viz., unusual dips (encircled in panel (b)),
a positive magnetization and an apparent asymmetry in $M(H)$ for
opposite signs of fields (panel (b), (d) and (f)) do not conform to
typical type\,-\,II superconducting behavior.}}
\end{figure}
The region close to $T_{c}$ has further been explored via the isothermal
($M(H)$) measurements. Figure~6 shows the $M$--$H$ responses at
$T=6.85$\,K (panels (a) and (b)), $6.90$\,K (panels (c) and (d))
and $6.95$\,K (panels (e) and (f)). The $M(H)$ curve at $T\approx7.1$\,K
(normal state) is also shown in panels (a), (c) and (e) for comparison.
In each case, the sample was first cooled in a field $H$ ($>H_{c2}(T)$)
from a temperature $T>T_{c}$ to the desired temperature. The magnetization
was then recorded while ramping the field to higher negative values
and then back to higher positive values. At $T\approx7.10$\,K, the
$M$--$H$ curve remains linear in the entire field range as expected
for the normal state ($T_{c}\approx7.1$\,K) of a superconductor.
Even though the $M$--$H$ curve at $T=6.85$\,K in Fig.~6(a) appears,
at first sight, to be typical of a type\,-\,II superconductor, some
anomaly can be clearly seen in the magnified view near $H=20$\,Oe
(see Fig.~6(b)). There are unusual dips (see encircled regions in
Fig.~6(b)) in magnetization near $H\approx0$, which are not expected
for a type\,-\,II superconductor. Moreover, there is a sharp positive
peak in the magnetization at $H\approx0$\,Oe and the magnetization
response is also found to be asymmetric about the $M$\,-\,axis,
as evident in Fig.~6(b). Note that the asymmetry in magnetization
about $M$\,-\,axis in the $M$--$H$ scan reminds of the asymmetry
observed earlier (see, for example, Fig.~3) in the temperature variations
of the field\,-\,cooled magnetization ($M_{FCC}(T)$) for the opposite
polarities of the magnetic field. At a higher temperature, $T=6.90$\,K
(cf. Fig.~6(c)), one can notice that the asymmetry in the $M$--$H$
curve persists, however, unlike the situation in Fig.~6(b), there
is no fingerprint of the dip like feature near $H\sim10$ Oe (see
the magnified view in Fig.~6(d)). The dips observed in Fig.~6(b)
have now morphed into a huge magnetization near zero field at $T=6.90$\,K
even though the system is still in the superconducting state. Similarly
at $T=6.95$\,K (Fig.~6(e)), the positive peak in magnetization
at $H\approx0$ and a nominal hysteresis in the $M$--$H$ curve (cf.
Fig.~6(f)) could be seen, which potray a completely different picture
than that for a usual type\,-\,II superconducting behavior.

\section{Discussion: $H$--$T$ phase space of surface superconductivity}

It is tempting to associate the anomalous features observed in Figs.~1
to 6, with some theoretical findings. The observation of paramagnetic
magnetization in the superconducting state of a conventional superconductor
is generally known to be an attribute of the compression of trapped
flux \cite{key-1,key-3,key-6} within the bulk of a sample that leads
to giant vortex states with multiple flux quanta ($L\gg1$). The self-consistent
solutions \cite{key-3,key-6,key-18} of the G\,-\,L equations for
certain geometries and various (mesoscopic) dimensions of superconducting
specimen yield multi\,-\,quanta states ($L\gg1$) at the onset of
surface superconductivity \cite{key-13}. These multi\,-\,quanta
states ($L\gg1$) tend to transform into pinned (single quantum, $L=1$)
Abrikosov lattice on lowering the temperature (in a constant field)
across the notional second critical field, $H_{c2}(T)$ line of the
$H$--$T$ space. However, such a transformation can follow a large
variety of paths \cite{key-6} involving several internal transitions
amongst the multi-quanta states. Each of these multi\,-\,quanta
states has a unique $M(H)$ loop associated with it such that the
magnetic response is diamagnetic at higher fields, which crosses over
to paramagnetic values at lower fields (see, for instance, Figs.~2
and 3 in Ref. \cite{key-6} and Figs.~16, 17 and 23 in Ref. \cite{key-18}).
If the superconducting state created (below $T_{c}$) gets stuck in
a certain $L$\,($>1$) state in a metastable manner, the magnetization
state initially ought to be diamagnetic (see Fig.~2(b) in Ref.~\cite{key-6}),
which then crosses over to paramagnetic values (along the $M(H)$
curve of a given $L$ state) as the temperature is swept down in a
constant field. The $M_{FCC}(T)$ curves below $3$\,mm amplitude
in Fig.~4(b) and those of Fig.~1 in a sense illustrate such a behaviour.

Since there is a specific field\,-\,temperature domain of each $L>1$
state \cite{key-6,key-18}, changes in the $H$ or $T$ values can
also bring transitions \cite{key-6} from one $L$ state into another
$L$ value. Such transitions are generally accompanied with abrupt
(jumps) changes in magnetization \cite{key-6,key-19}. The undulations
overriding the PME and the oscillatory response in Figs.~1 to 5 could
be viewed as elucidation of such transitions. According to Zharkov\textquoteright s
results \cite{key-6}, the magnetization signal will remain diamagnetic
if transitions continue to happen involving the neighbouring states
following the lowest free energy criterion (see Fig.~2(b) in Ref.
\cite{key-6} for transitions between $L$ and $L\pm1$ states). However,
if one allows for metastability and for transitions to happen between
different metastable states, it can lead to a complex magnetization
response such that, magnetization signal can change from a given paramagnetic/diamagnetic
value to higher/lower value (see Fig.~3(b) in Ref. \cite{key-6}). 

Features emanating from the $M_{FCC}(T)$ curves of Fig.~4(b) elucidate
a wide variety, from diamagnetic to paramagnetic values (induced by
the change in vibration amplitude) in the $M_{FCC}(T)$ responses
below $T_{c}$, which can be associated with the metastable nature
arising out of a variety in the multi\,-\,quanta states in the domain
of surface superconductivity. The non\,-\,uniqueness in magnetization
is also well apparent from Fig.~3 wherein the $M_{FCC}(T)$ is recorded
at constant amplitude ($=0.5$\,mm) for opposite polarities of a
given $H$. Since the non\,-\,uniqueness in $M_{FCC}(T)$ persists
even for lower temperatures as the saturated $M_{FCC}(T)$ values
are significantly different in the inset panels of Figs.~3(a), 3(b)
and Fig.~4(b), one may also argue that a non\,-\,unique coexistence
of multi\,-\,quanta ($L>1$) and single quantum ($L=1$) states
is being witnessed far below $T_{c}$. 

Another interesting facet of Zharkov\textquoteright s findings \cite{key-6}
derives from the free energy plots of various $L$ states (see Fig.~3(c)
in Ref. \cite{key-6}), which predicts a short temperature window
just below $T_{c}$, where the single quantum (Abrikosov) state does
not exist. Consistent with this prediction, the $M(H)$ loops recorded
in the proximity of $T_{c}$ (cf. Fig.~6) in Ca$_{\mbox{3}}$Ir$_{4}$Sn$_{13}$
display anomalous features which are not anticipated for a pinned
type\,-\,II superconductor. The lack of evidence for (pinned) Abrikosov
vortex state ($L=1$) above $6.95$\,K in Ca$_{3}$Ir$_{4}$Sn$_{13}$
presumably echo the predictions of Zharkov \cite{key-6}.

\begin{figure}[!t]
\begin{centering}
\includegraphics[scale=0.38]{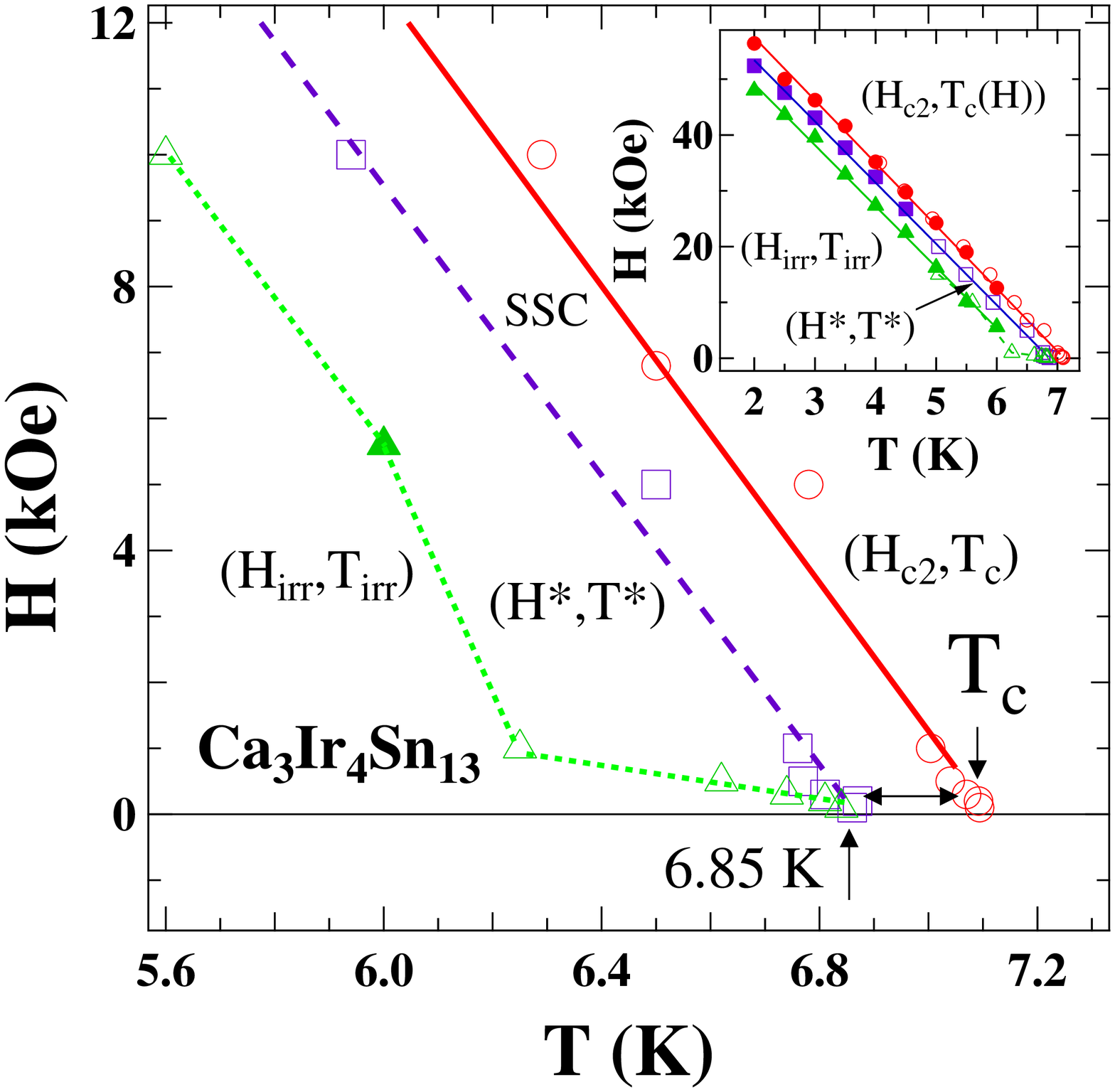}
\par\end{centering}

\protect\caption{{\footnotesize{}(Color online) A portion of the $H$--$T$ phase diagram
for Ca$_{3}$Ir$_{4}$Sn$_{13}$ (extracted from our earlier report,
Ref. \cite{key-14}). It comprises an irreversibility ($H_{irr}$,$T_{irr}$)
line, the characteristic field/temperature ($H^{*}$,$T^{*}$) line
and the upper critical field ($H_{c2}$,$T_{c}$) line as obtained
from Fig.~4. The inset in Fig.~6 displays the complete phase diagram.
The phase\,-\,space region bounded by ($H^{*}$,$T^{*}$) and ($H_{c2}$,$T_{c}$)
lines has been surmised to be the region of surface superconductivity.}}
\end{figure}
From the viewpoint of surface superconductivity, we are now inclined
to reconsider the field\,-\,temperature ($H$--$T$) phase\,-\,space
of Ca$_{3}$Ir$_{4}$Sn$_{13}$ reported in our earlier investigation
\cite{key-14}, which is reproduced in Fig.~7. The ($H^{*}$\,,\,$T^{*}$)
line, as obtained from Fig.~5, in this phase diagram can now be surmised
as a crossover regime between the bulk and the surface superconductivity.
The ($H_{c2}$,$T_{c}(H)$) line in Fig.~7 can be surmised as the
third critical field line. The region enclosed between the ($H^{*}$,$T^{*}$)
and ($H_{c2}$,$T_{c}(H)$) lines is identified as the region where
surface superconductivity (SSC) dominates. As stated above, the $H$--$T$
region bounded by the interval $6.85$ K\,$<T<T_{c}$ does not imbibe
fingerprints of Abrikosov flux line lattice; it is only below $T\approx6.85$\,K,
the mean\,-\,field description is expected to be applicable in Ca$_{3}$Ir$_{4}$Sn$_{13}$.
In this context, we believe that the ($H^{*}$,$T^{*}$) line meeting
the $T$\,-\,axis at $T\approx6.85$\,K (surmised as bulk superconducting
transition) is reasonable. Across the interval $T^{*}<T<T_{c}$, the
$M(T)$ (Fig.~5) and the $M$--$H$ (Fig.~6) responses reflect the
metastable nature of the multi\,-\,quanta ($L>1$) vortex states
occurring in the domain of surface superconductivity.

\section{Summary}

We have investigated a weakly-pinned single crystal of a low $T_{c}$
superconducting compound, Ca$_{3}$Ir$_{4}$Sn$_{13}$ via dc magnetization
measurements. Recently we explored \cite{key-14} the pinning behaviour,
order\,-\,disorder transitions in the vortex matter and the vortex
phase diagram in this compound. The said vortex phase diagram of Ca$_{3}$Ir$_{4}$Sn$_{13}$
in our previous report \cite{key-14} had left some issues unresolved,
for example, a non\,-\,linear behaviour in the reversible magetization
response ($M(T)$/$M(H)$) was observed close to the $T_{c}(H)$ line.
The low field $M_{FCC}(T)$ data in the present report reveals some
intriguing features, viz., anomalous paramagnetic signals, a rich
multiplicity or non\,-\,uniqueness in magnetization ranging from
diamagnetic to paramagnetic values, oscillatory magnetization response
below $T_{c}$, etc. These novel features have been associated with
the occurrence of multi\,-\,quanta states in the domain of surface
superconductivity and the metastability effects associated due to
the transitions amongst various multi\,-\,quanta states ($L>1$)
in this compound. The isothermal $M$--$H$ responses in the temperature
range $6.85$\,K$<T<T_{c}$ (Fig.~6), corroborate further the notion
of multi\,-\,quanta ($L>1$) vortex states in the realm of surface
superconductivity. 

The paramagnetic magnetization ($M_{FCC}(T)$) observed in Ca$_{3}$Ir$_{4}$Sn$_{13}$
is ascribed to the multi\,-\,quanta states ($L>1$) occurring in
the domain of surface superconductivity, even though, other explanations
behind the paramagnetic nature cannot be ruled out. About a decade
ago, such a behavior was theoretically anticipated using the G\,-\,L
theory for specific situations \cite{key-3,key-6,key-18}. We believe
that the present work has provided some experimental evidences rationalizing
the predictions of those theoretical works \cite{key-3,key-6,key-18}.
\begin{acknowledgments}
\noindent Santosh Kumar would like to thank the Council of Scientific
and Industrial Research, India for grant of the Senior Research Fellowship.\end{acknowledgments}

\end{document}